\documentclass{article}

\usepackage{arxiv}

\usepackage[utf8]{inputenc} 
\usepackage[T1]{fontenc}    
\usepackage{hyperref}       
\usepackage{url}            
\usepackage{booktabs}       
\usepackage{amsfonts}       
\usepackage{nicefrac}       
\usepackage{microtype}      
\usepackage{lipsum}
\usepackage{amsmath}
\usepackage{graphicx}
\usepackage{booktabs,caption}
\usepackage[flushleft]{threeparttable}

\title{Multinuclear Absolute MR Thermometry}

\author{
  Emilia V. Silletta \\
  New York University \\
  Department of Chemistry \\
  New York, NY 10003, USA;\\
  Universidad Nacional de Córdoba \\
  Facultad de Matemática Astronomía, Física y Computación \\
  Medina Allende s/n, X5000HUA, Córdoba, Argentina; \\
  Instituto de Física Enrique Gaviola \\
  CONICET, Medina Allende s/n, X5000HUA, Córdoba, Argentina \\
  \texttt{emiliasilletta@unc.edu.ar} \\
   \AND
   Alexej Jerschow \\
   New York University \\
   Department of Chemistry \\New York, NY 10003, USA;\\
  \texttt{alexej.jerschow@nyu.edu} \\
   \AND
   Guillaume Madelin\textsuperscript{*}\\
   New York University School of Medicine \\
   Department of Radiology \\
   Center for Biomedical Imaging \\
   New York, NY 10016, USA \\
   \texttt{Guillaume.Madelin@nyulangone.org} \\
    \AND
    Leeor Alon\thanks{Shared last authorship}\\
    New York University School of Medicine \\
    Department of Radiology \\
    Center for Biomedical Imaging \\
    New York, NY 10016, USA \\
    \texttt{Leeor.Alon@nyulangone.org} \\
   }
\begin{document}
\maketitle

\begin{abstract}
Non-invasive measurement of absolute temperature is important for proper characterization of various pathologies and for evaluation of thermal dose during interventional procedures. The proton magnetic resonance (MR) frequency shift method can be used to map relative temperature changes in tissues; however, spatiotemporal variations in the main magnetic field and the lack of an internal frequency reference at each location challenge the determination of absolute temperature. Here, we introduce a novel multinuclear method for absolute MR thermometry, based on the fact that the proton and sodium nuclei exhibit a unique and distinct characteristic frequency dependence with temperature. A one-to-one mapping between the precession frequency difference of the two nuclei and absolute temperature is demonstrated. Proof-of-concept experiments were conducted in aqueous solutions with different NaCl concentrations, in agarose gel samples, and in freshly excised ex vivo mouse tissues. One-dimensional chemical shift imaging was also conducted at two steady-state temperatures, demonstrating excellent agreement with infrared measurements.
\end{abstract}

\keywords{Nuclear Magnetic Resonance \and Absolute Temperature Mapping \and Thermometry \and Sodium Frequency Shift \and Proton Frequency Shift}

\section{Introduction}
Magnetic resonance imaging (MRI) has become a valuable diagnostic tool for visualization of subtle pathologies with millimeter resolution. In recent years there has been a growing interest in the utilization of MR techniques to measure temperature changes in vivo \cite{Reike2008}. While most MR contrast mechanisms vary with temperature change \cite{bloembergen1948relaxation, simpson1958diffusion, nelson1987temperature, hall1990observation, dickinson1986measurement,bloembergen1948relaxation, bottomley1984review,dickinson1986measurement, delannoy1991noninvasive}, it has been shown that the proton resonance frequency (PRF) method has greatest sensitivity to thermal change in most tissues \cite{hindman1966proton}. The temperature dependence of the PRF was first discovered by Hindman when conducting NMR experiments on intermolecular forces and hydrogen bond formation \cite{hindman1966proton}, and adapted to estimate temperature change through MR phase imaging measurements by Ishihara et al. \cite{ishihara1995precise} and De Poorter et al. \cite{poorter1995noninvasive}. The method is currently the gold standard for mapping thermal changes in interventional applications such as high-intensity focused ultrasound (HIFU) \cite{holbrook2010real, roujol2012robust, kickhefel2010accuracy}, radiofrequency (RF) hyperthermia \cite{delannoy1990hyperthermia}, RF ablation \cite{vandenbosch2008}, and RF power deposition from wireless devices \cite{alon2015method}. 

The PRF method relies on the subtraction of pre- and post-exposure phase images, or on the local determination of the frequency shift of protons with MR spectroscopy (MRS), to calculate temperature change due to exposure conditions \cite{poorter1995noninvasive2, young1996evaluation}. However, non-thermal $B_0$ changes, such as due to movement \cite{rieke2007referenceless, vigen2003triggered}, magnet field drift \cite{peters2000magnetic}, flow \cite{peters2000magnetic}, or shim changes, greatly limit the applicability of the PRF method. Today, PRF thermometry is restricted to experiments with large thermal gradients or phantom studies with minimal $B_0$ drift throughout the experiment. Furthermore, PRF methodologies are not capable of reconstruction of absolute temperature in tissues, because an internal frequency reference (in each voxel) is required. Knowledge of the absolute temperature in tissues is particularly important due to the correlation of many pathologies with thermal disruption and is fundamental for quantification of thermal dose during interventional procedures \cite{vanrhoon2016,vanRhoon2013,LeavittP2016,Wang2014,TANG2008312,Zaretsky2018}. 

In nuclear magnetic resonance (NMR) experiments, internally-referenced measurements of temperature are widely used to monitor temperature of samples by measuring the chemical shift between two or more temperature-dependent peaks such as between the OH and CH\textsubscript{2} groups in ethylene glycol \cite{Raiford1979}. Internally-referenced experiments are robust against instabilities of $B_0$ because changes in macroscopic $B_0$ equally shift the independent peaks \cite{van1968calibration, raiford1979calibration}, enabling the reconstruction of absolute temperature. In the brain, the amid proton in N-acetylaspartate (NAA) peak has been utilized as a temperature-independent reference. However, due to the low concentration of NAA in the brain ($\sim$10 mM) \cite{rigotti2011longitudinal}, challenges associated with water suppression, pH-dependent separation of the NAA-water peaks, and imaging time required to obtain adequate signal-to-noise ratio (SNR), absolute thermometry via imaging of the NAA peak remains challenging \cite{dehkharghani2015proton}. Overall, utilization of a frequency reference, endogenous to most tissues, for the reconstruction of absolute temperature has not been attainable since the advent of NMR.

In this work, we introduce a novel multinuclear approach for absolute MR thermometry. We demonstrate that \textsuperscript{23}Na nuclei exhibit an NMR frequency shift dependency with temperature that is roughly twice that of the \textsuperscript{1}H nuclei. Thus, measuring the difference of NMR frequencies of the \textsuperscript{23}Na and \textsuperscript{1}H nuclei provides a one-to-one mapping with temperature, allowing absolute temperature reconstruction that is robust against macroscopic $B_0$ inhomogeneities. Proof-of-concept experiments were conducted in aqueous solutions with different NaCl concentrations, in agarose gel samples, and in freshly-excised ex vivo mouse tissues. One-dimensional chemical shift imaging (CSI) was also performed for two steady-state temperature regimes.

\section{Theory}
\subsection*{Temperature dependence of the NMR frequency shift} 
The Larmor frequency $f^N$ of the magnetic moment of a nucleus $N$ is determined by the magnetic field $B_{nuc}$ that the nucleus experiences and the gyromagnetic ratio $\gamma^N$ of the nucleus. $B_{nuc}$ is the result from a shielding constant $\sigma^N$ altering the macroscopic magnetic field $B_0$ according to:
\begin{equation}
    f^N=\frac{\gamma}{2\pi}B_{nuc}=\frac{\gamma^N}{2\pi}(1-\sigma^N)B_{0}.
    \label{eq:freq}
\end{equation}
The shielding constant is expressed as: 
\begin{equation}
    \sigma^N=\sigma^N_{i}+\sigma^N_{\chi}+\sigma^N_{e}
    \label{eq:sigma}
\end{equation}
where $\sigma^N_{i}$ is the intramolecular shielding constant, $\sigma^N_{e}$ is the intermolecular electric shielding effect, and  $\sigma^N_{\chi}$ is the volume magnetic susceptibility shielding effect of nucleus $N$. Both $\sigma^N_{\chi}$ and $\sigma^N_{e}$ can change with temperature $T$. The precession frequency can thus be expressed as:
\begin{equation}
    f^N(T) = \frac{\gamma^N}{2\pi}\Big[1-\sigma^N_i-\sigma^N_\chi(T)-\sigma^N_e(T)\Big]B_0.
    \label{eq:fX}
\end{equation}
By defining $ f^N_0 = \frac{\gamma^N}{2\pi}B_0 $, we can calculate the frequency shift $\delta f^N(T)$ of a nucleus $N$, in parts-per-million (ppm), as:
\begin{equation}
    \delta f^N(T) = \frac{f^N_0-f^N(T)}{f^N_0}.
    \label{eq:dfN_ppm_def}
\end{equation}
This can be expressed as the sum of a temperature-independent component and a temperature-dependent component:
\begin{equation}
    \delta f^N(T) = \Big[\sigma^N_i\Big] + \Big[\sigma^N_\chi(T) + \sigma^N_e(T)\Big].
    \label{eq:dfN_ppm_calc}
\end{equation}
Since the temperature dependency of $\sigma^N_\chi$ and $\sigma^N_e$ is linear with temperature \cite{hindman1966proton,hindman1962nuclear}, the susceptibility and electric shielding can be written: 
\begin{subequations}
    \begin{align}
    \sigma^N_\chi(T) & = \sigma^N_{\chi0} + \alpha^N_\chi \cdot T, \label{eq:sigma_chi_e_Ta} \\
    \sigma^N_e(T) & = \sigma^N_{e0} + \alpha^N_e \cdot T. \label{eq:sigma_chi_e_Tb}
    \end{align}
\end{subequations}

Equations \ref{eq:sigma_chi_e_Ta} and \ref{eq:sigma_chi_e_Tb} can be combined such that the nucleus' frequency shift is rewritten as a constant $\sigma^N_0$ (in ppm) and a frequency shift thermal coefficient $\alpha^N$ (in ppm/$^{\circ}$C):
\begin{equation}
    \delta f^N(T) = \sigma^N_0 + \alpha^N \cdot T,
    \label{eq:dfN_ppm_T}
\end{equation}
with
\begin{subequations}
    \begin{align*}
    \sigma^N_0 & = \sigma^N_i+\sigma^N_{\chi0}+\sigma^N_{e0} \\
    \alpha^N & = \alpha^N_\chi + \alpha^N_e.
    \end{align*}
    \label{eq:sigma0_alpha}
\end{subequations}

\subsection*{Measurement of relative temperature change} 

The frequency shift thermal coefficient $\alpha^N$ can be calibrated for a specific nucleus (e.g. \textsuperscript{1}H) and a sample of interest. Since $\delta f^N$ can vary with local $B_0$ fluctuations (shim, motion, field drift), and the component $\sigma^N_0$ is generally unknown and can vary due to different electronic and susceptibility shieldings, absolute temperature cannot be calculated using Eq. \ref{eq:dfN_ppm_T}. This equation can however be used to measure relative temperature changes using nucleus $N$ = \textsuperscript{1}H MRS or MRI (PRF method) and a calibrated value $\alpha^N\sim$ -0.01 ppm/$^{\circ}$C \cite{hindman1962nuclear} in human tissues. By subtracting the frequency shifts measured at 2 different times (e.g. before and after heating), the effect of $\sigma^N_0$ is cancelled and relative temperature changes are calculated as:
\begin{equation}
    \Delta T = T_1-T_2 = \frac{\delta f^N(T_1)-\delta f^N(T_2)}{\alpha^N}.
    \label{eq:DeltaT_PRF} 
\end{equation}

\subsection*{Measurement of absolute temperature} 
Absolute temperature can be derived from Eq. \ref{eq:dfN_ppm_T} by detecting the frequency shift of two nuclei within the same sample or voxel (in case of localized MRS or MRI), where the difference between their respective frequency shift thermal coefficients $\alpha$ and constants $\sigma_0$ are well-known theoretically or calibrated experimentally. Using the following definitions for two nuclei $N \equiv A,B$ (which can even be of the same species, but from a different molecule or local environment): 
\begin{subequations}
    \begin{align}
    \Delta f(T) & = \delta f^A(T) - \delta f^B(T), \\
    \Delta \sigma_0 & = \sigma^A_0 - \sigma^B_o, \\
    \Delta \alpha & = \alpha^A - \alpha^B \neq 0,
    \end{align}
    \label{eq:delta_definitions1}
\end{subequations} 
the frequency shift difference between the two nuclei can thus be written:
\begin{equation}
    \Delta f(T) = \Delta \sigma_0 + \Delta \alpha \cdot T.
    \label{eq:delta_f_ppm}
\end{equation}
Upon calibration of $\Delta \sigma_0$ and $\Delta \alpha$ for the two nuclei and samples of interest (fluid, tissue), absolute temperature of the sample can be calculated as follows:
\begin{equation}
    T = \frac{\Delta f(T)-\Delta \sigma_0}{\Delta\alpha}.
    \label{eq:calculate_T}
\end{equation}

\section*{Methods}


\subsection*{NMR experiments}
Experiments were carried out on an 11.7 T NMR Bruker Avance I spectrometer (Bruker BioSpin) operating at 500.19 MHz for \textsuperscript{1}H, and 132.3 MHz for \textsuperscript{23}Na, using a 5 mm double resonance broadband probe. The test tubes with different samples under investigation (aqueous solutions with different NaCl concentrations, agarose gel, ex vivo tissues) were placed inside the spectrometer where the sample temperature could be controlled using gas flow and a temperature sensor providing a precise, stable and reliable temperature regulation. After each desired temperature was reached, a standard free induction decay (FID) pulse sequence was used with a 90\textsuperscript{$\circ$} pulse. The duration of the pulse is 11 $\mu$s and 9 $\mu$s for \textsuperscript{1}H and \textsuperscript{23}Na, respectively, and 8 averages were used with TR = 15 s for \textsuperscript{1}H, and 0.5 s for \textsuperscript{23}Na, dwell time = 100 $\mu$s, spectral width = 5 kHz. 

\subsection*{Solutions and agarose gel}
Solution samples with 11 different NaCl concentrations ($C$ = 0.1, 1, 2, 5, 8, 11, 14, 17, 20, 23, 26 \% weight) were prepared by mixing $x$ mg of NaCl in $(10-x)$ mg of deionized water in a beaker (with $x$ = 0.01, 0.1, 0.2, 0.5, 0.8, 1.1, 1.4, 1.7, 2.0, 2.3, 2.6), and transfered to 5 mm NMR tubes. The solution at 26\% weight correspond to NaCl saturation in water. Corresponding NaCl concentrations in mol/L can be calculated according to the equations \ref{eq:rho_calibration} and \ref{eq:C_calibration} below, and results are presented in Table \ref{tab:nacl_concentration}. A gel was prepared by mixing 2\% w/v of agarose with 1\% w/v NaCl in deionized water. The gel mixture was incrementally heated in a microwave to fully dissolve the agarose. The solution was poured into a 5 mm NMR tube forming a uniform, homogeneous gel upon cooling.

\subsection*{Tissue samples}
Four tissues samples (brain, kidney, liver and muscle) were obtained from two female mice whose weights were 22.2 and 25 g. 

\subsection*{Effect of pH}
In order to study the effect of pH on the multinuclear MR temperature measurements, solutions with different pH values were tested for $\Delta\alpha$ and $\Delta\sigma_0$ calibration. The solutions of different pH values were prepared by adding a small amount of acid HCl or base KOH solutions to the water solution sample with 1\% weight NaCl, to adjust to the desired pH value. The pH was measured with a Fisher Scientific\textsuperscript{PM} accumet\textsuperscript{PM} AB150 pH Benchtop Meter and calibrated with three standard buffers with pH values 4.01, 7 and 10.01. The reported pH values were measured before acquiring the NMR data. The pH range was from 4.9 to 9.07. The results are shown in Figure \ref{fig:ph} and demonstrate that pH has negligible influence on the $\Delta\alpha$ and $\Delta\sigma_0$ values.

\subsection*{Heating system and 1D CSI procedure}
An in-house built alternating-current resistive heating setup was constructed to create an NMR-compatible heating setup that does not interfere with the multinuclear NMR acquisition \cite{Gilchrist2016anmri}. A signal generator (B071HJ31WN, KKmoon, China), operating at 100 KHz was connected to a 130W class D amplifier (TPA3250D2EVM, Texas instruments Inc., USA). The output of the amplifier was connected an in-house built low pass filter with a cutoff frequency of 10 MHz to mitigate radiofrequency waves being picked up and transmitted in close proximity to the RF coil in the NMR spectrometer. The output of the low pass filter was connected to a resistive wire insert made of wounded AWG 32G enameled copper wire (ECW32AWG1LB, Bntechgo Inc., USA) placed inside the 5 mm NMR test tube filled with 2\% agarose and 1\% NaCl in water. A baseline proton 1D chemical shift imaging (CSI) acquisition was conducted with the following imaging parameters: 16 steps in the z-encoding, 1 average, and a repetition time of 15 s, giving a total experimental time of 5 min. A sodium 1D CSI acquisition over the same field of view was then acquired with the following parameters: 16 steps in the z-encoding, 32 averages, and a repetition time of 0.3 s, with a total experimental time of 5 min. The 1D CSI pulse sequence consisted of a 90\textsuperscript{$\circ$} pulse followed by a pulse gradient which encodes the spatial position in z-direction. 

\begin{figure}      %
    \centering
    \includegraphics[width=0.550\linewidth]{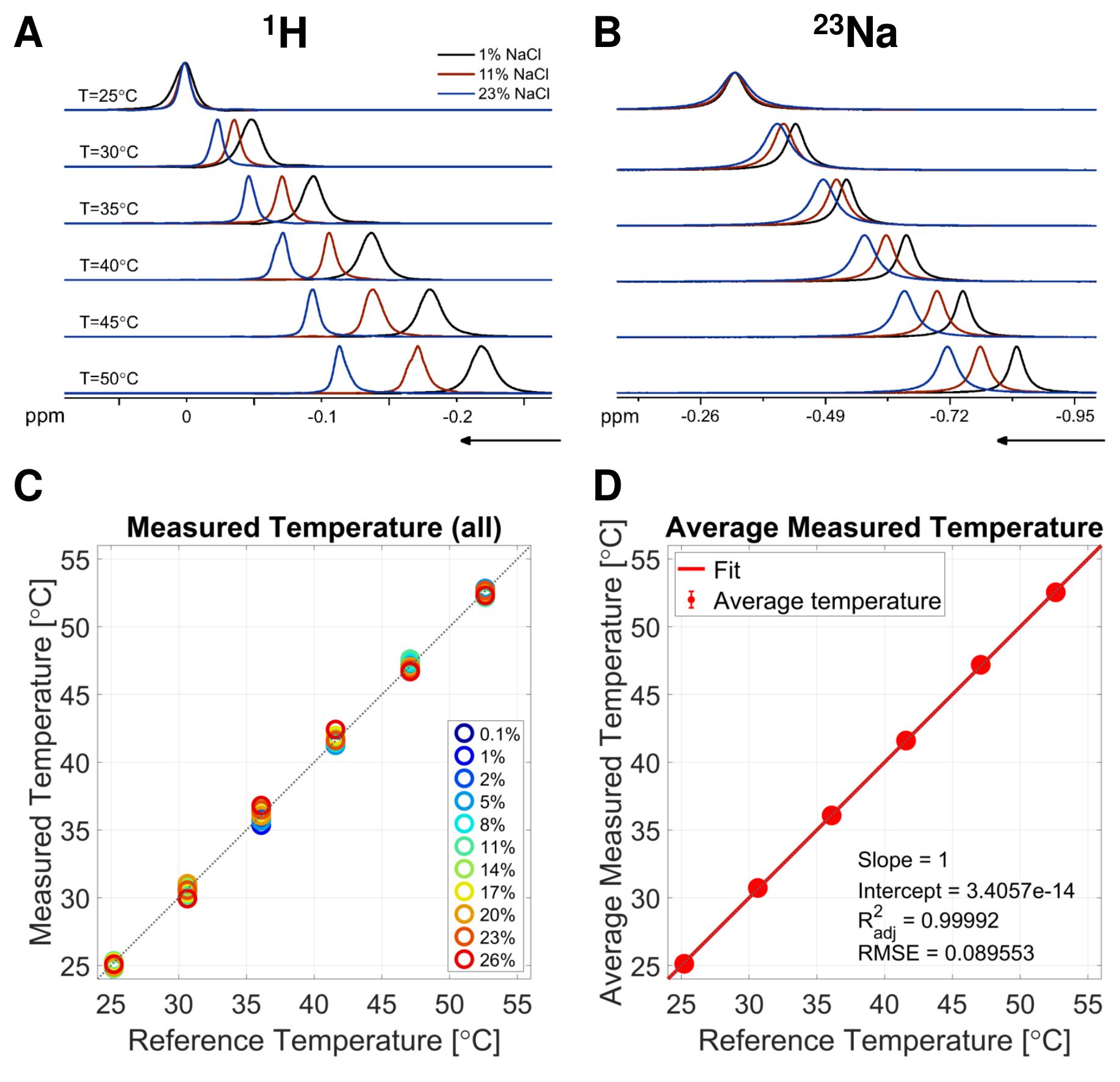}
    \caption{Examples of (\textbf{A}) \textsuperscript{1}H and (\textbf{B}) \textsuperscript{23}Na spectra at 6 different temperatures for 3 solutions with different NaCl concentrations: 1\%, 11\% and 23\% weight. (\textbf{C}) Temperature measurements in all the calibration samples containing 0.1\% to 26\% NaCl, using $\Delta\alpha$ and $\Delta\sigma_0$ calculated from the results shown in Figure \ref{fig:delta_alpha_sigma}. (\textbf{D}) Average temperature measurements over all samples.}
    \label{fig:spectra_calibration_temperature_solutions}
\end{figure}
After the baseline proton and sodium acquisitions were conducted, a 1V peak-to-peak sinusoidal waveform was used to drive the amplifier. The waveform at 100 KHz was used in order to not interfere with the RF, gradient or $B_0$ field. Sample temperature was monitored in real time with the internal temperature probe of the Bruker 500MHz spectrometer to ensure that heating of the sample was in a steady state. After twenty minutes, a steady state of the temperature was attained, and CSI acquisitions were acquired at proton and sodium frequencies. Sodium and proton spectra were then used to reconstruct the absolute temperature. The absolute temperature was plotted and compared with infrared temperature measurements acquired at steady state temperature using a FLAIR infrared (IR) camera (E75, FLIR Systems Inc., USA).  

\section*{Results}

\begin{table}       
    \centering
    \caption{$\Delta\alpha$ and $\Delta\sigma_0$ for different samples.} 
    \begin{tabular}{llll}
    Samples & Calibration & $\Delta\alpha$ (ppm/\textsuperscript{$\circ$}C) & $\Delta\sigma_0$ (ppm) \\
    \midrule
    Solution 1\%    & Solutions 0.1-26\%         & 0.010690  & 0.056744 \\
    Solution 1\%    & Self-calibration           & 0.010301  & 0.070416 \\
    Solution 0.3\%  & Solutions 0.1-26\% (fit)   & 0.010720  & 0.075141 \\
    Agarose         & Self-calibration           & 0.010321  & 0.075775 \\
    Brain           & Self-calibration           & 0.011060  & 0.023918 \\
    Muscle          & Self-calibration           & 0.011887  & -0.01556 \\
    Kidney          & Self-calibration           & 0.011206  & 0.029343 \\
    Liver           & Self-calibration           & 0.011500  & 0.034852 \\
    \bottomrule
    \end{tabular}
    \label{tab:delta_alpha_sigma0}
\end{table}

All experiments were performed at 11.7 T with the following spectrometer reference frequencies: $f^{H}_0$ = 500.2031765 MHz, $f^{Na}_0$ = 132.3120951 MHz (fixed ratio $f^{H}_0$/$f^{Na}_0$ = 3.7804796). We first measured $\Delta\alpha$ and $\Delta\sigma_0$ in 11 samples with NaCl concentrations ranging from 0.1\% to 26\% (saturation) by weight. For each solution, NMR spectra were acquired at 6 different temperatures, as measured by the spectrometer sensor: 25, 30, 35, 40, 45 and 50\textsuperscript{$\circ$}C (the corresponding real temperatures corrected using the spectrometer temperature calibration are shown in SI: Materials and Methods, and Table S1). The position of the peak maximum followed a linear trend with temperature, with the slope corresponding to the frequency shift thermal coefficient $\alpha$, and the intercept $\sigma_0$. 

\begin{figure}  
    \centering
    \includegraphics[width=0.550\linewidth]{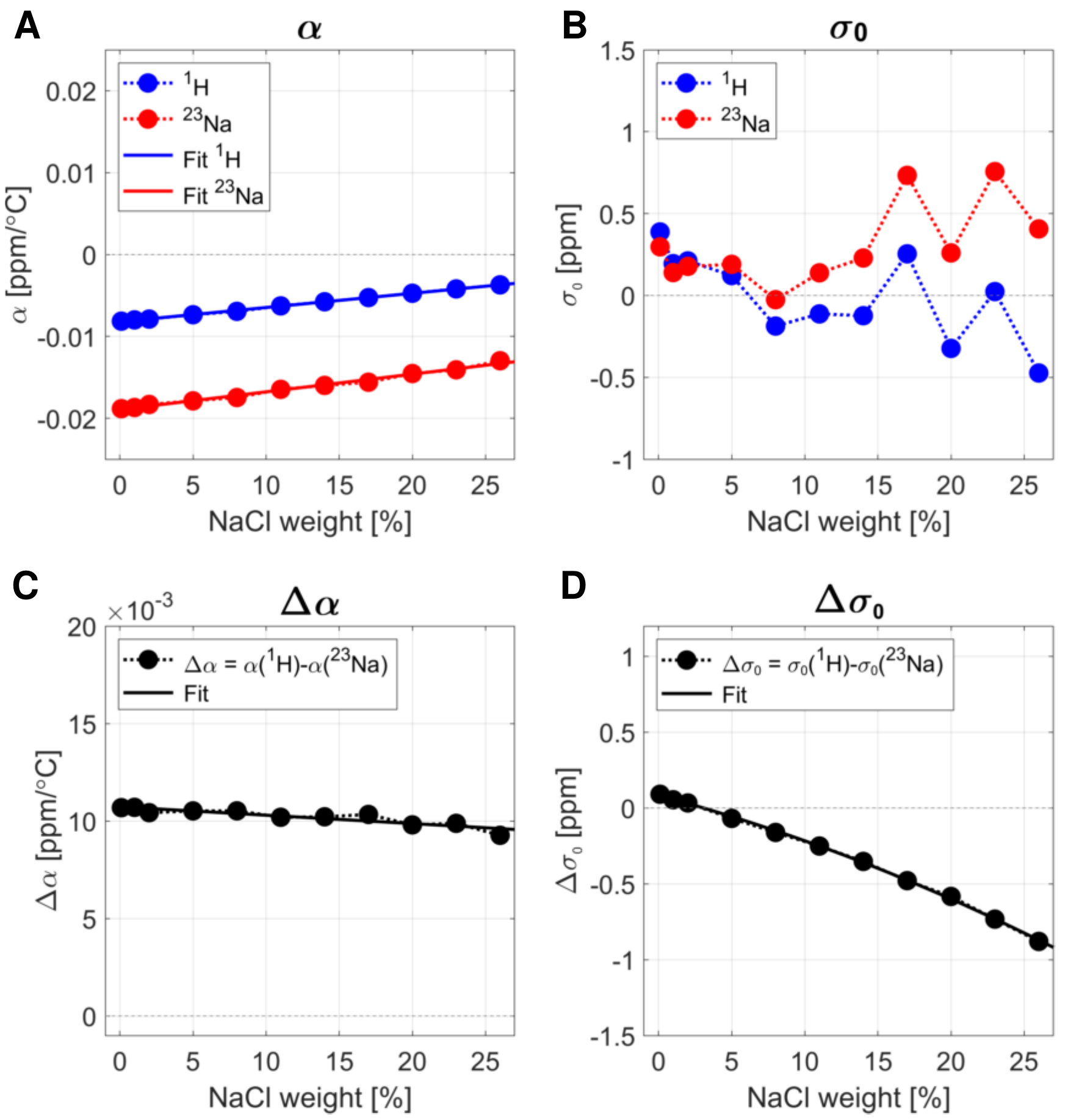}
    \caption{(\textbf{A}) \textsuperscript{1}H and \textsuperscript{23}Na NMR frequency shift thermal coefficients $\alpha$ (slope of the linear fit) at different NaCl concentrations. (\textbf{B}) \textsuperscript{1}H and \textsuperscript{23}Na constants $\sigma$\textsubscript{0} (intercept of the linear fit) at different NaCl concentrations. (\textbf{C}) $\Delta\alpha$ calculated from (\textbf{A}). (\textbf{D}) $\Delta\sigma$\textsubscript{0} calculated from (\textbf{B}).}
    \label{fig:delta_alpha_sigma}
\end{figure}

Figure \ref{fig:spectra_calibration_temperature_solutions}(A,B) shows examples of \textsuperscript{1}H and \textsuperscript{23}Na spectra at different temperatures, where the frequency changes with temperature are shown to vary with the NaCl concentration. Figure \ref{fig:spectra_calibration_temperature_solutions}(C) shows the measured temperature for all NaCl solutions after calibration of $\Delta\alpha$ and $\Delta\sigma_0$ from the results shown in Figure \ref{fig:delta_alpha_sigma}, compared to the reference temperature at which the experiments were performed. The average measured temperature for all solutions is plotted in Figure \ref{fig:spectra_calibration_temperature_solutions}(D), showing excellent agreement with reference temperature (adjusted $R^2_{adj}$ = 0.99992, and root mean square error RMSE = 0.09$^{\circ}$C). 

Figure \ref{fig:delta_alpha_sigma} shows the results of the linear fitting of the frequency shift of \textsuperscript{1}H and \textsuperscript{23}Na versus temperature for the 11 samples. The fits are shown in SI, Figures S1-S11. The frequency shift thermal coefficient $\alpha$ for \textsuperscript{1}H, shown in Figure \ref{fig:delta_alpha_sigma}(A), is consistent with literature, where the value of approximately -0.01 ppm/\textsuperscript{$\circ$}C is typically found for low NaCl concentrations (1\% weight or less in biological tissues). It was found that the frequency shift thermal coefficient $\alpha$ for \textsuperscript{23}Na was approximately twice higher in magnitude than for \textsuperscript{1}H. The \textsuperscript{1}H and \textsuperscript{23}Na spectra for each sample were acquired on the same day at 6 temperatures, and the same shim was used for both nuclei. Different samples were acquired on different days in the following order: 1\%, 26\%, 11\%, 17\%, 23\%, 8\%, 0.1\%, 2\%, 5\%, 14\%, 20\%. This random order ensures that the smooth variation that was detected for $\Delta\alpha$ and $\Delta\sigma_0$ with NaCl concentrations was not an effect of the spectrometer magnetic field drift or $B_0$ shim changes on different days. As shown on Figure \ref{fig:delta_alpha_sigma}(C,D), both $\Delta\alpha$ and $\Delta\sigma_0$ showed a smooth variation with NaCl concentration, even when individual $\sigma_0$ values for \textsuperscript{1}H and \textsuperscript{23}Na seem to fluctuate randomly in different samples acquired on different days. The variation of $\Delta\alpha$ is linear with increased NaCl concentration, while the variation of $\Delta\sigma_0$ shows a nearly linear decrease with increasing NaCl concentration. 

\begin{figure}
\centering
\includegraphics[width=0.550\textwidth]{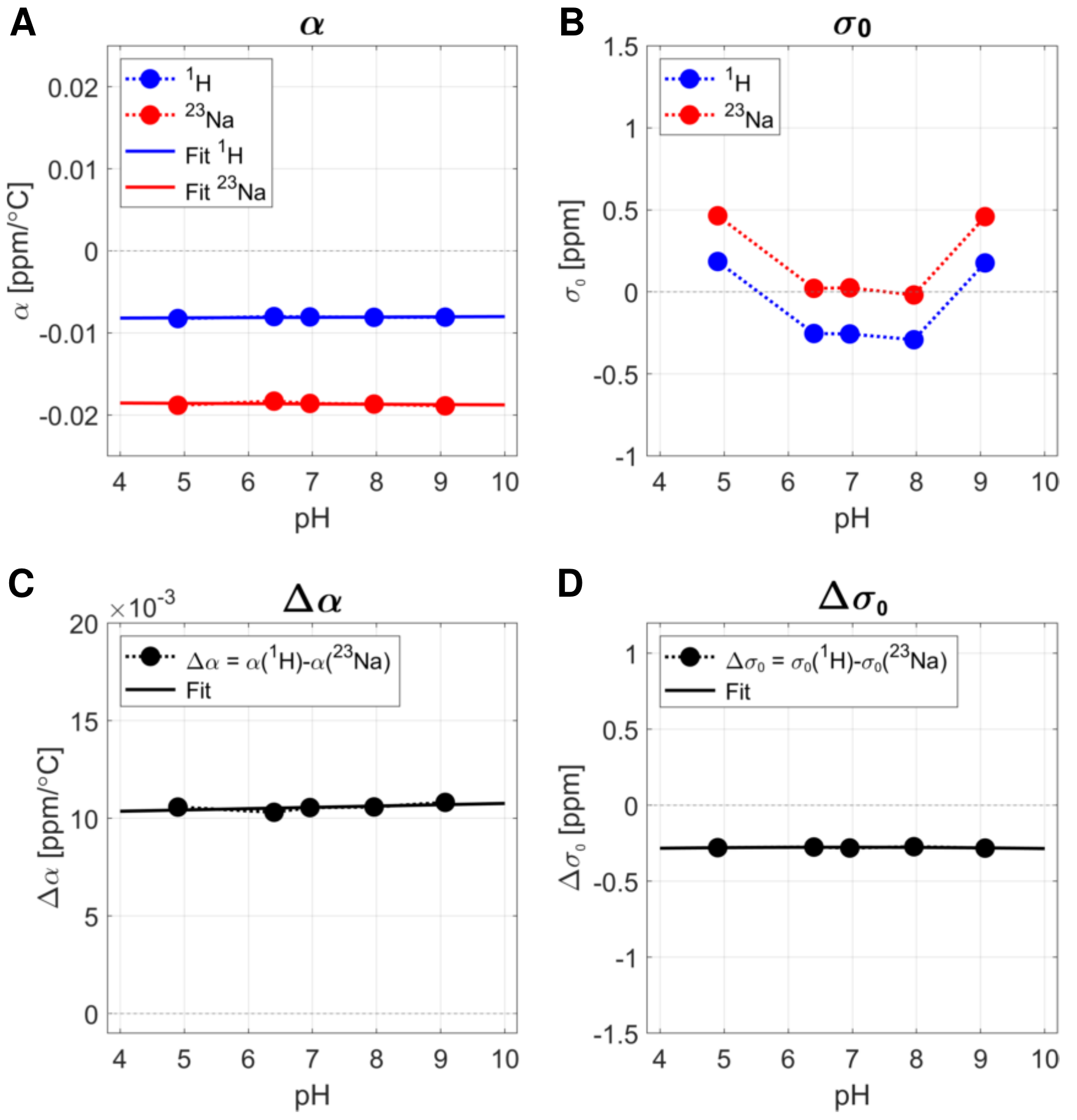}
\caption{$\alpha$, $\sigma_0$, $\Delta\alpha$ and $\Delta\sigma_0$ values for \textsuperscript{1}H and \textsuperscript{23}Na in a solution with 1\% NaCl, for different pH.}
\label{fig:ph}
\end{figure}

In order to study the effect of pH on the multinuclear MR temperature measurements, solutions with different pH values were tested for $\Delta\alpha$ and $\Delta\sigma_0$ calibration. The pH range was from 4.9 to 9.07. The results are shown in \ref{fig:ph} and demonstrate that pH has negligible influence on the $\Delta\alpha$ and $\Delta\sigma_0$ values.

\begin{figure}  
    \centering
    \includegraphics[width=0.550\linewidth]{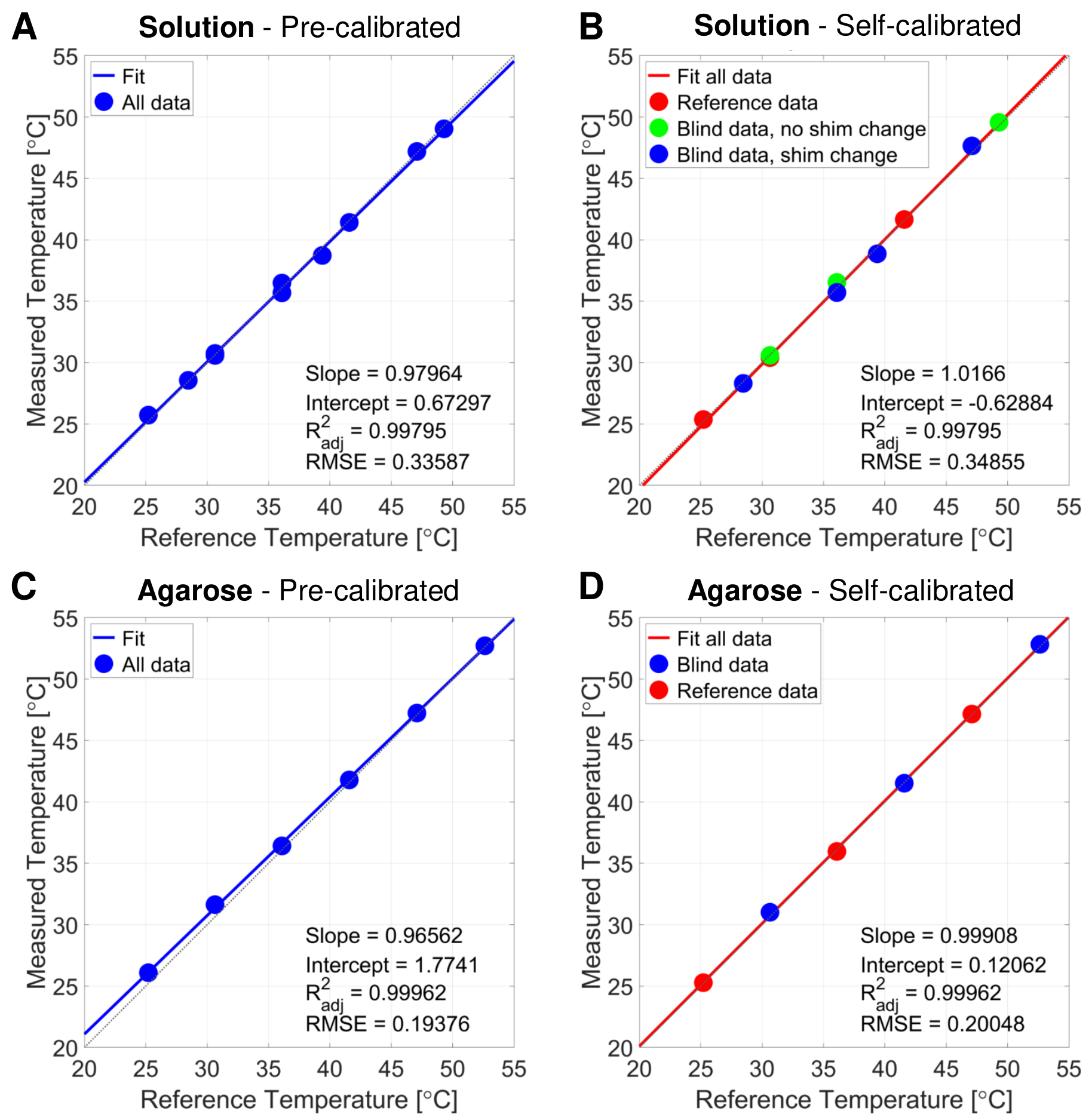}
    \caption{(\textbf{A}) Absolute temperature measurements using the pre-calibrated $\Delta\alpha$ and $\Delta\sigma_0$ from the 1\% NaCl solution. (\textbf{B}) The first 3 experiments were used to self-calibrate $\Delta\alpha$ and $\Delta\sigma_0$ (red dots), then 3 temperatures were reconstructed from blind experiments at random temperatures with no shim changes (green dots), and the last 4 experiments included both blind temperatures and shim changes (blue dots), all in the 1\% NaCl solution. (\textbf{C}) Temperature measurements for the 2\% agarose sample with 1\% NaCl using the $\Delta\alpha$ and $\Delta\sigma_0$ calibration from the 1\% NaCl solution. (\textbf{D}) Measured temperatures using 3 frequency measurements to self-calibrate $\Delta\alpha$ and $\Delta\sigma_0$ (red dots) in the agarose sample, and 3 blind data (blue dots).}
    \label{fig:temp_meas_solution_agarose}
\end{figure}

In order to test the ability of the method to predict unknown temperatures, 10 experiments were then carried out on a solution with NaCl concentration of 1\% weight (similar to physiological conditions). Figure \ref{fig:temp_meas_solution_agarose}(A) shows the calculated temperatures for all the data using the $\Delta\alpha$ and $\Delta\sigma_0$ calibration obtained with the 1\% solution used in Figure \ref{fig:delta_alpha_sigma} and Table \ref{tab:delta_alpha_sigma0}. As a next step, three peak frequency measurements at 25, 30 and 40$^{\circ}$C were used to self-calibrate $\Delta\alpha$ and $\Delta\sigma_0$ for this sample, plotted by red dots in Figure \ref{fig:temp_meas_solution_agarose}(B). The sample was then brought to three random blind temperatures with the same shimming conditions (green dots in Figure \ref{fig:temp_meas_solution_agarose}(B)). Then, the sample was brought to four more random blind temperatures where the magnet shims were randomly changed to alter $B_0$ (blue dots in Figure \ref{fig:temp_meas_solution_agarose}(B)). All calculated temperatures in \ref{fig:temp_meas_solution_agarose}(A,B) were in excellent agreement with the reference temperatures ($R^2_{adj}$ = 0.998, RMSE $\sim0.34^{\circ}$C). A similar experiment was conducted in a sample with 2\% agarose and 1\% NaCl. Figure \ref{fig:temp_meas_solution_agarose}(C) shows the results of the calculated temperature plotted against the reference value using the pre-calibrated $\Delta\alpha$ and $\Delta\sigma_0$ from the 1\% NaCl solution used in Figure \ref{fig:spectra_calibration_temperature_solutions}(C). In Figure \ref{fig:temp_meas_solution_agarose}(D), three frequency measurements were used to self-calibrate $\Delta\alpha$ and $\Delta\sigma_0$ in the gel itself, and three blind temperatures were calculated. In both cases, pre-calibration in a 1\% NaCl solution and self-calibration in gel lead to very similar results with accurate and precise measurement of the sample temperatures ($R^2_{adj}$ = 0.999, RMSE $\sim0.20^{\circ}$C). 

\begin{figure}  
    \centering
    \includegraphics[width=1\linewidth]{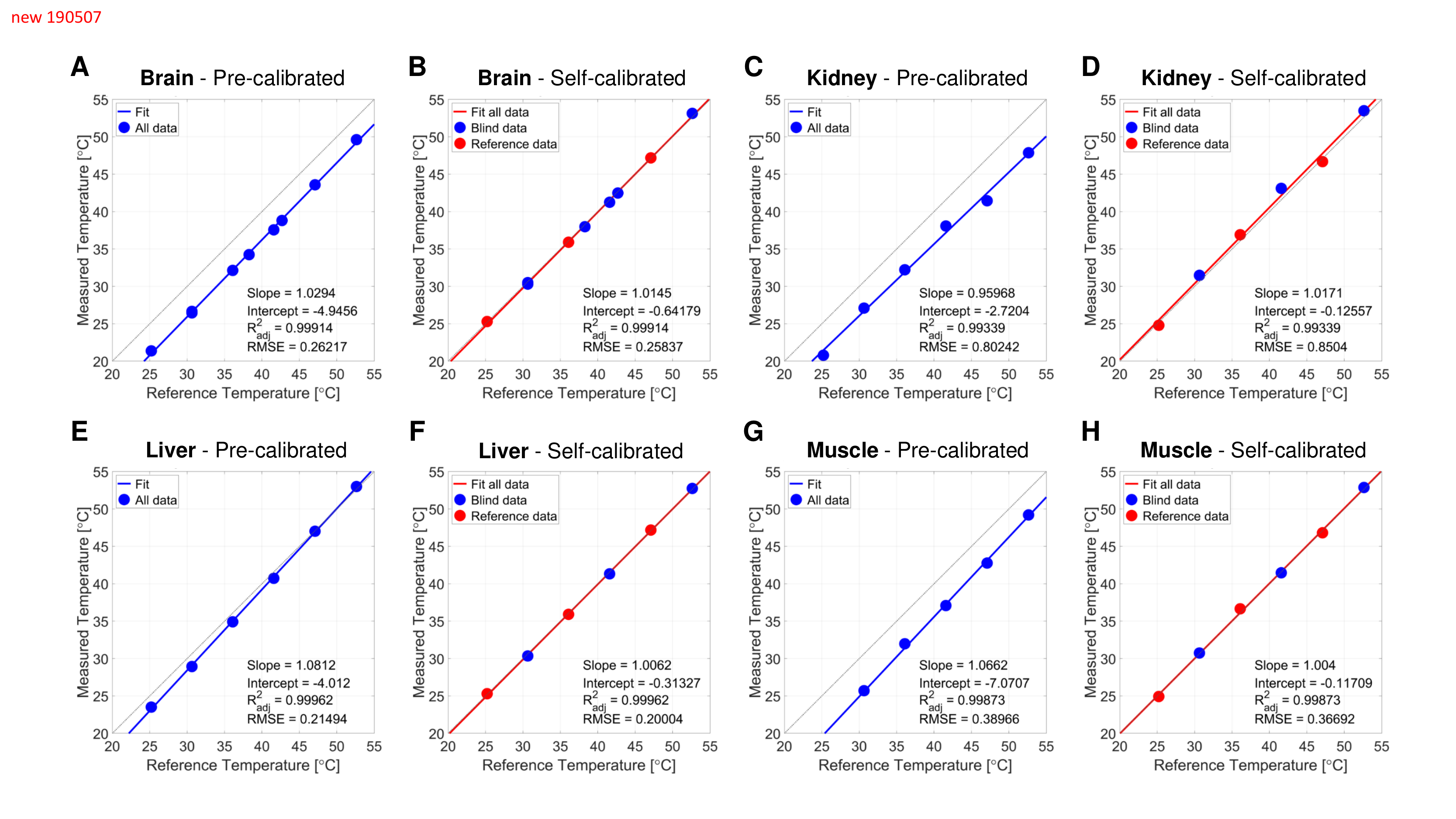}
    \caption{Absolute temperature measurements in ex vivo mouse tissue samples (brain, kidney, liver, muscle), using either pre-calibration of $\Delta\alpha$ and $\Delta\sigma_0$ from 0.3\% NaCl (fitted values) or self-calibration of $\Delta\alpha$ and $\Delta\sigma_0$ using 3 known temperatures (25, 35 and 45$^{\circ}$C, red dots) and blind measurements (blue dots). (\textbf{A}) Brain, pre-calibrated. (\textbf{B}) Brain, self-calibrated. (\textbf{C}) Kidney, pre-calibrated. (\textbf{D}) Kidney, self-calibrated. (\textbf{E}) Liver, pre-calibrated. (\textbf{F}) Liver, self-calibrated. (\textbf{G}) Muscle, pre-calibrated. (\textbf{H}) Muscle, self-calibrated. }
    \label{fig:temp_meas_tissues_calib_self}
\end{figure}

Figure \ref{fig:temp_meas_tissues_calib_self} shows the temperatures measured in freshly excised ex vivo mouse tissues (brain, kidney, liver, and muscle) using three peak frequency measurements at 25, 35 and 45$^{\circ}$C for self-calibrating $\Delta\alpha$ and $\Delta\sigma_0$, and other blind temperatures were calculated from this self-calibration. In all tissues, an excellent agreement was found for the calculated temperature when this self-calibration procedure was used. However, when the pre-calibration of $\Delta\alpha$ and $\Delta\sigma_0$ was calculated from a 0.3\% NaCl solution (or approximately 50 mM, similar to biological tissue concentrations) from fitting of the data measured at 0.1 to 26\% NaCl, a constant offset is detected. Pre-calibration of $\Delta\alpha$ and $\Delta\sigma_0$ from the 1\% NaCl and the 0.1\% NaCl solution were also tested, with similar results than with 0.3\% NaCl. Only in the case of liver, the pre-calibrated temperature measurement showed a good agreement with the reference temperature, as shown in Figure \ref{fig:temp_meas_tissues_calib_self}(E). The main difference in sample preparation was that the consistency of the liver sample was still homogeneous when introduced in the NMR tube, while the other tissue samples were composed of small pieces, leading to a more inhomogeneous system which increased the susceptibility effects significantly (air bubbles, fat mixture within the tissue), resulting in a constant temperature offset.

\begin{figure}
    \centering
    \includegraphics[width=1\linewidth]{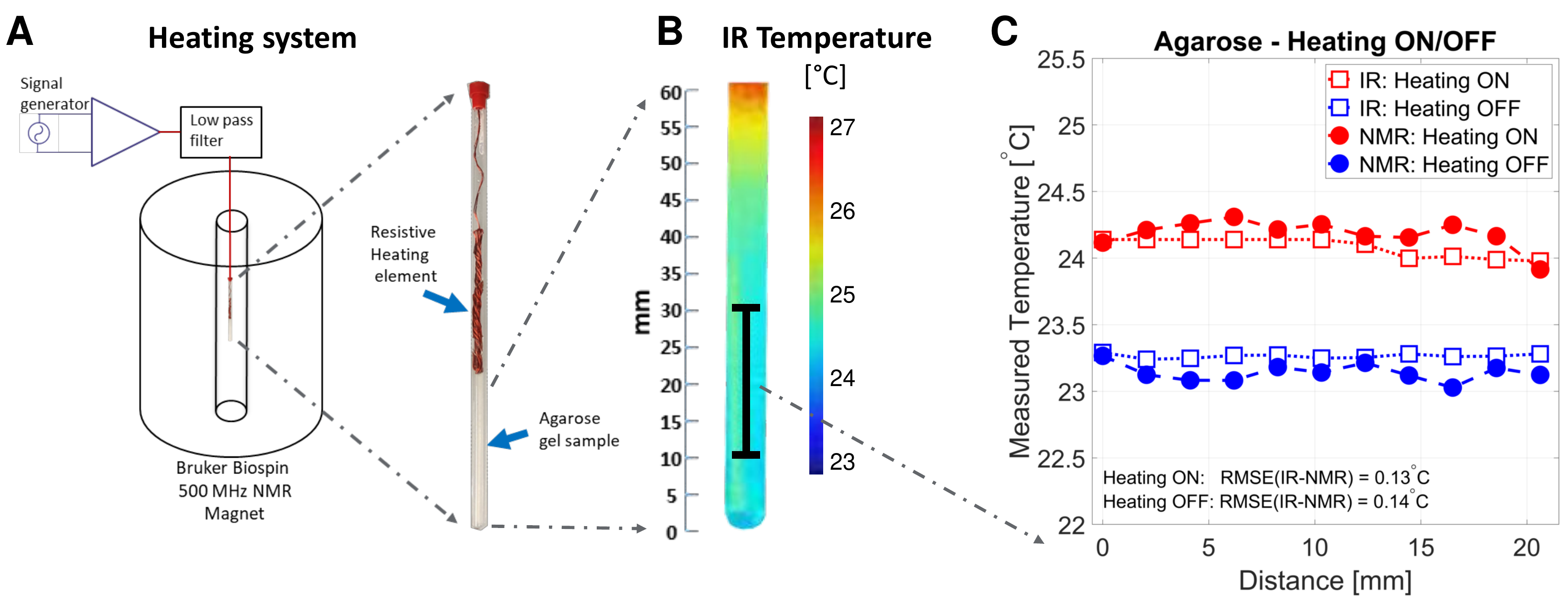}
    \caption{Comparison of 1D CSI results with infrared (IR) imaging. (\textbf{A}) Heating system setup. (\textbf{B}) Steady state temperature profile measured using IR in the distal 60 mm section of the NMR tube. The black line represents the 1D imaging volume probed using the absolute thermometry method. (\textbf{C}) Absolute temperature reconstructed when heating was OFF and in steady state. The root mean square error (RMSE) of the difference in temperature measurements between the IR and NMR measurements was 0.13\textsuperscript{$\circ$}C and 0.14\textsuperscript{$\circ$}C, for the heating ON and OFF conditions, respectively.}
    \label{fig:csi_agarose}
\end{figure}

Finally, in order to test the ability to map absolute temperature spatially, a 1D CSI measurement was carried out as shown in Figure \ref{fig:csi_agarose}. The experiment was conducted in the gel sample with 2\% agarose and 1\% NaCl. The heating system setup is shown in Figure \ref{fig:csi_agarose}(A) and the spatial temperature map of the sample measured with an infrared (IR) camera is shown in Figure \ref{fig:csi_agarose}(B). Figure \ref{fig:csi_agarose}(C) shows the measured temperatures using both IR camera (open square) and CSI data (closed circles) over 20 mm in the sample (NMR-detectable zone) before and after heating the sample. The measured temperatures using both methods are in good agreement, showing an increase of 1\textsuperscript{$\circ$}C along the entire sample after sample heating. 

The values of $\Delta\alpha$ and $\Delta\sigma_0$ used in this paper for calculating temperatures in 1\% NaCl solution, in agarose gel, and in tissue samples, are summarized in Table \ref{tab:delta_alpha_sigma0}.

\section*{Discussion}

In this work, we present a new method of multinuclear absolute MR thermometry which takes advantage of the different and unique frequency shifts of the sodium and proton nuclei with temperature. The method is validated in fluid samples with different NaCl concentrations, in agarose gels, and in ex vivo fresh tissue from mice, with precise temperature control. The method was shown to be robust to $B_0$ inhomogeneities which are a challenge for thermometry methods such as the PRF.

Changes in water proton frequency shifts with temperature reflect changes in the hydrogen-bonded structure of water \cite{schneider1958proton, hindman1966proton, muller1965concerning, ruterjans1966chemical}. The nature of these changes has been studied extensively, and two main models have been proposed to explain it \cite{hindman1966proton, pople1951molecular}. In the first model, the temperature-induced frequency shift of water relates to the stretching and bending of the hydrogen bonds which are responsible for the electrical shielding effect \cite{pople1951molecular}. The second model describes a change in electrical shielding due to the breaking of the hydrogen bonds. Specifically, a steady state is created between ice-like lattice water structure, where hydrogen bonds are fully formed, and a monomeric water structure where no hydrogen bonds are present. These two models, when used independently, cannot fully explain the temperature- and ionic concentration-dependent frequency shift of water. Consequently, a mixed model where hydrogen bond length stretching and bending (model 1) alongside hydrogen bond rearrangements (model 2) best explains and predicts experimental results on the temperature and ionic concentration dependency of the water frequency shift \cite{hindman1968proton}. The effect of strong electrolytes (such as NaCl) causes a concentration-dependent shift in the proton resonance frequency, with some electrolytes inducing an increase in the frequency, while others, such as Na\textsuperscript{+}, inducing a reduction in the frequency \cite{wertz1956nuclear, hindman1962nuclear, shoolery1955nuclear}. The chloride ion Cl\textsuperscript{-} has been shown to have a small effect on the proton frequency shift relative to that of Na\textsuperscript{+} \cite{hindman1962nuclear}. When the sodium ion is surrounded by water, a hydration shell is created, where, depending on the temperature, four to eight \cite{ohtaki1992structural, mancinelli2007hydration, chizhik1987nmr} molecules of water can temporarily coordinate a single Na\textsuperscript{+}. In such solutions, water molecules can be in an unbound state with the ion (free water outside the hydration shell), which causes minimal change to the electrostatic structure of the hydrogen bond. For a fraction of time, water molecules are in a bound state with the ion (hydration shell) \cite{malinowski1966nmr}, causing a structural modification to the hydrogen bond, thus altering the electrical shielding of the \textsuperscript{1}H nucleus. The time for which water is bound to the ion is dependent on the NaCl concentration. 

With respect to the frequency shift of the sodium ion, a strong correlation with the frequency shift of water was observed, suggesting that a temperature-related modification of the hydrogen bonds coexists with a modification of the electrical shielding of the sodium ion. A temperature rise increases the effective hydrogen bond length of water, increasing the negative charge distribution around the oxygen nucleus within the water molecule. This increase in negative charge distribution intensifies the ion-dipolar attraction between oxygen and sodium, consequently enhancing the electrical shielding of the sodium nucleus. As the concentration of NaCl increases, the magnitude of the frequency shift thermal coefficient $\alpha$ of sodium decreases due to the competition between the ions for the water molecules, causing a decrease in average time for which water is bound to the ion \cite{hindman1968proton}. These effects form the basis for the multinuclear absolute thermometry method, enabling a sample-specific bijective mapping between the frequency difference of \textsuperscript{1}H and \textsuperscript{23}Na nuclei and temperature.

Our results demonstrate that once the proposed multinuclear thermometry method was calibrated on the aqueous solution with 1\% NaCl, the frequency shift difference between the \textsuperscript{1}H and \textsuperscript{23}Na nuclei can be used to calculate the absolute temperature of the same sample under different shimming conditions with high accuracy (with an error of the order of 0.3\textsuperscript{$\circ$}C for temperatures between 25 and 50\textsuperscript{$\circ$}C). When calibration of the multinuclear thermometry method was conducted in aqueous solutions and then applied to predict the temperature in ex vivo tissue samples (brain, muscle, liver and kidney), a constant temperature offset was observed. We believe that this offset can occur due to two main factors influencing the calibration of $\Delta\alpha$ and $\Delta\sigma_0$: (1) the preparation of the tissue samples, and (2) the presence of multiple ions inside the tissue samples. In case (1), the tissue samples were inserted in small pieces into the 5 mm NMR tubes, thus creating relatively inhomogeneous samples with air bubbles that are artificially inducing strong local susceptibility effects which are significantly stronger than under in vivo conditions. This tissue susceptibility was not present in the aqueous solution calibration of $\Delta\alpha$ and $\Delta\sigma_0$, and is most likely the main source of error. An exception was the liver sample that was kept uniform and homogeneous in the tube, hence a closer agreement between the pre-calibrated and the self-calibrated temperature measurements was found. In case (2), previous studies have shown that, for example, the presence of potassium ions K\textsuperscript{+} can cause a proton frequency shift, while other ions generally induce smaller shifts due to their small chemical shift effect or their smaller concentrations in tissues \cite{hindman1962nuclear, shoolery1955nuclear}. These ions were not present in the liquid samples, yet present in tissues at varying concentrations. The effect of these ions on the sodium resonance frequency shift is poorly understood and needs further future investigation. 

Studies have shown that the volume of magnetic susceptibility changes linearly with temperature \cite{peters1999heat}, and its effect on the \textsuperscript{1}H resonance frequency shift is roughly an order of magnitude smaller than the electrical shielding effect \cite{peters1999heat, odeen2019magnetic, poorter1995noninvasive2}. As a result, calibration of the absolute thermometry method on the sample includes the sample-specific magnetic susceptibility shielding information for both sodium and proton. While susceptibility changes are accounted for in the model, measurement of temperature in voxels with very high susceptibility that alters the lineshapes of the spectra can be challenging since the reconstruction relies on detection of the proton and sodium spectra's center frequency. This effect was observed in our CSI measurements, where voxels close to the edge of the tube and close to the metallic resistive heating apparatus had to be excluded from the reconstruction due to spectral distortion. 

In conclusion, we present a proof-of-concept method for measuring the absolute temperature non-invasively in samples using a multinuclear magnetic resonance approach, based on the detection of the frequency shift difference between two different nuclei (\textsuperscript{1}H and \textsuperscript{23}Na is this case). This absolute temperature mapping method is compatible with an implementation using phase MRI or localized MRS at both the \textsuperscript{1}H and \textsuperscript{23}Na frequencies, for potential medical applications. Phase measurement acquisitions are more time-efficient than spectroscopic imaging as long repetition times needed to obtain high spectral resolution are not necessary, which can have an impact on the timing of clinical scanning. Translation and optimization of the multinuclear absolute thermometry technique to in vivo imaging, where both sodium and proton phase images can be acquired simultaneously or in an interleaved fashion \cite{LEE1986343}, will be the subject of a future investigation.

\section*{Acknowledgement}
The authors want to thank Dr Seena Dehkharghani for useful discussions on the mechanisms underlying the proton and sodium frequency shift dependence with temperature. This work was supported in part by the National Institutes of Health (NIH): grants R01EB026456, R01NS097494, R21CA213169 and P41EB017183. The work was also supported by an award from the US National Science Foundation: NSF CBET 1804723.

\begin{table}\centering
\caption{Spectrometer temperature calibration.}
\begin{tabular}{lllllll}
Temperature (\textsuperscript{$\circ$}C) \\
\midrule
Theoretical (spectrometer) & 25 & 30 & 35 & 40 & 45 & 50 \\
Corrected (real) & 25.21 & 30.64 & 36.09 & 41.58 & 47.09 & 52.63 \\
\bottomrule
\label{tab:temperature_calibration}
\end{tabular}
\end{table}

\bibliographystyle{unsrt}  
\bibliography{references}

\newpage
\section*{Supplemental Material}

\subsection*{NaCl concentration calibration}
The conversion of NaCl concentration $C$ from \% weight (\%wt) unit to mol/L unit was calculated using the two following equations:
\begin{equation*}
    \rho = a\cdot C_{\%wt}^{2}+b\cdot C_{\%wt}+c,
    \label{eq:rho_calibration}
\end{equation*}
with the density of water $\rho$ in kg/L or g/mL, a = 1.682$\times10^{-5}$, b = 0.007079 and c = 0.9984, calculated using the data from the CRC Handbook of Chemistry and Physics (86th ed), p. 8-71 \cite{lide2005crc}, on the properties of water-NaCl mixtures (density of water at different NaCl concentrations in \%wt); and
\begin{equation*}
    C_{mol/L} = \frac{C_{\%wt}}{100}\times\frac{\rho}{M}\times1000,
    \label{eq:C_calibration}
\end{equation*}
with M = 58.44 g/mol the molar mass of NaCl.

\subsection*{Spectrometer temperature calibration}
The sample temperature was controlled with a variable temperature system, which is part of the Bruker spectrometer. The gas flow streams through a pipe along the sample tube and leaves the probe head at the top. A temperature sensor measures the temperature and gives the value to a control unit that regulates the heater power to keep the temperature constant. Since the temperature sensor is not inside the NMR tube, a calibration must be done in a sample with a known temperature-dependence behavior. Calibration data was previously acquired on this spectrometer on methanol, where the chemical shift difference between the peaks correlates to the real temperature \cite{ammann1982simple}. The range of temperatures was from -50\textsuperscript{$\circ$}C to 67\textsuperscript{$\circ$}C. The real temperatures calculated from the frequency shifts of the peaks are compared to the temperatures obtained from the control unit in the spectrometer, and the following fitting parameters were obtained between the real temperature $T_{real}$ in the sample and the temperature measured by the spectrometer sensor $T_{spec}$: 
\begin{equation*}
    T_{real} = a\cdot T^{2}_{spec}+b\cdot T_{spec}+c,
    \label{eq:T_calibration}
\end{equation*}
with a = 5.944181$\times10^{-4}$, b = 1.052388, and c = -1.470807. We used the same correction for all of our experiments. See Table \ref{tab:temperature_calibration} for the resulting corrected (real) temperatures used in our experiments. The frequency shifts of the \textsuperscript{1}H and \textsuperscript{23}Na signals were detected at each temperature by tracking the position of the maximum of peak of their NMR spectrum. All data processing was performed in Matlab (The MathWorks Inc., Natick, MA, USA).

\subsection*{Measurements of frequency shift thermal coefficient $\alpha$ and constant $\sigma_0$ for \textsuperscript{1}H and \textsuperscript{23}Na in solutions}
The frequency shift thermal coefficient $\alpha$ and constant intercept $\sigma_0$ were measured in 11 solutions with different NaCl concentrations ($C$ = 0.1, 1, 2, 5, 8, 11, 14, 17, 20, 23, 26 \% weight), by fitting the frequency shift $f$ of the maximum of the NMR peak versus 6 temperatures ($T$ = 25, 30, 35, 40, 45, 50$^{\circ}$C), for both the \textsuperscript{1}H and \textsuperscript{23}Na nuclei:
\begin{equation*}
    f = \alpha T + \sigma_0.
    \label{eq:fit_freq_shift_temperature}
\end{equation*}

\subsection*{Fitting of $\alpha$, $\sigma_0$, $\Delta\alpha$, $\Delta\sigma_0$ versus NaCl concentrations}
The values of $\alpha$, $\sigma_0$, $\Delta\alpha$, $\Delta\sigma_0$ measured at different NaCl concentrations $C_{\%wt}$ were fitted using the equations below. The fitting parameters are given in Table \ref{tab:fit_parameters_alpha_sigma0}. \\

For $\alpha$ and $\sigma_0$ of \textsuperscript{1}H and \textsuperscript{23}Na, and for $\Delta\alpha$:
\begin{subequations}
    \begin{align*}
    \alpha = a\cdot C_{\%wt}+b. \\
    \Delta\alpha= a\cdot C_{\%wt}+b.
    \end{align*}
    \label{eq:fit_alpha_delta_alpha}
\end{subequations}

For $\Delta\sigma_0$:
\begin{equation*}
    \Delta\sigma_0 = a\cdot C_{\%wt}^{2}+b\cdot C_{\%wt}+c.
    \label{eq:fit_delta_sigma0}
\end{equation*}

\begin{table}\centering
\caption{NaCl concentration calibration}
\begin{tabular}{llllllllllll}
[NaCl] \\
\midrule
in \% weight & 0.1 & 1 & 2 & 5 & 8 & 11 & 14 & 17 & 20 & 23 & 26 \\
in mol/L    & 0.017 & 0.172 & 0.346 & 0.885 & 1.446 & 2.030 & 2.637 & 3.269 & 3.924 & 4.605 &	5.311 \\
\bottomrule
\label{tab:nacl_concentration}
\end{tabular}
\end{table}

\begin{table}
\begin{threeparttable}
    \centering
    \caption{Fitting parameters for $\alpha$, $\sigma_0$, $\Delta\alpha$, $\Delta\sigma_0$  versus NaCl concentrations in weight \% ($C_{\%wt}$), corresponding to Eq. \ref{eq:fit_delta_sigma0} and \ref{eq:fit_alpha_delta_alpha}.} 
    \begin{tabular}{lllll}
    Fit parameters & $\alpha$ for \textsuperscript{1}H & $\alpha$ for \textsuperscript{23}Na &	$\Delta\alpha$ & $\Delta\sigma_0$ \\
    \midrule
    a & 0.000172 & 0.000215 & -4.303e-05 & -0.0004218\\
    95\% CB for a & (0.000168,0.000177) & (0.000199,0.000232) & (-5.735$\times10^{-5}$,-2.87$\times10^{-5}$) & (-0.0005425,-0.000301)\\
    b & -0.008183 &	-0.01892 & 0.01073 & -0.02573\\
    95\% CB for b & (-0.008248,-0.008117) & (-0.01915,-0.01868) & (0.01053,0.01094) & (-0.02884,-0.02262)\\
    c & & & & 0.0829\\
    95\% CB for c & & & & (0.06756,0.09823)\\
    R\textsuperscript{2}\textsubscript{adj} & 0.99866 &	0.98882 & 0.81874 & 0.99898\\
    RMSE & 5.77$\times10^{-5}$ & 2.09$\times10^{-4}$ & 1.83$\times10^{-4}$ & 0.01056\\
    \bottomrule
    \end{tabular}
    
    \begin{tablenotes}
      \small
      \item 
      \sffamily Abbreviations: CB = Confidence Bounds; RMSE = Root Mean Square Error; R\textsuperscript{2}\textsubscript{adj} = adjusted R\textsuperscript{2}.
    \label{tab:fit_parameters_alpha_sigma0}
    \end{tablenotes}
    
    \end{threeparttable}
\end{table}

\end{document}